\documentclass[a4paper,twoside,openany,11pt]{memoir} 
\pdfoutput=1

\title{
    Symmetry Restoration in the SMEFT: Finite Counterterms for a Non-Anticommuting gamma-5
}
\author{
    J. Fuentes-Martín, A. Moreno-Sánchez, and A.E. Thomsen
}

\def\fullheadfoot{0} 
\usepackage[english]{babel}
\usepackage{microtype,xspace}
\usepackage[utf8]{inputenc}	

\usepackage{amsfonts,amsmath,amssymb,bm,mathrsfs,mathtools,dsfont} 	
\allowdisplaybreaks 	
\usepackage[table]{xcolor}
\usepackage{hyperref}
\usepackage{slashed}
\usepackage{multicol}

\usepackage{siunitx}
\usepackage{geometry}
\usepackage[sort&compress,numbers,merge]{natbib}
\usepackage{chapterbib}
\addto\captionsenglish{\renewcommand*{\bibname}{References}}
\makeatletter
\renewcommand{\@memb@bchap}{ 
\bibmark \prebibhook
}
\makeatother

\usepackage{graphicx}
\graphicspath{{./Figures/}}
\usepackage{caption,subcaption}
\usepackage{multirow}

\usepackage{tabularx}
\newcolumntype{Y}{>{\centering\arraybackslash}X}

\usepackage[shortlabels]{enumitem}
\setlist{itemsep=.1em,topsep=.5em}
\SetEnumerateShortLabel{i}{\textit{\roman*}}

\definecolor{red}{rgb}{0.6,.0706,.1373}
\definecolor{blue}{rgb}{0,0.396,0.741}
\definecolor{green}{rgb}{0.25,0.6,0.2}
\definecolor{teal}{rgb}{0.11,0.6,0.6}
\definecolor{orange}{rgb}{.8, .4806, 0.173}
\definecolor{yellow}{rgb}{.8, .7, 0.05}
\colorlet{blueref}{blue!80!black}
\colorlet{bluelink}{blue!90!black}
\hypersetup{
	colorlinks, 
	bookmarksopen, 
	bookmarksnumbered,
	citecolor=blueref, 		
	linkcolor=bluelink,	
	urlcolor=bluelink,			
	pdftitle={\thetitle},    
	pdfauthor={\theauthor},    	
}

\addto\captionsenglish{}

\renewcommand{\contentsname}{Contents}
\renewcommand{\printtoctitle}[1]{}

\settocdepth{subsection}
\newcommand{\toc}{ {
	\hypersetup{linkcolor = black} 
	\vspace*{-.06\textheight}	
	\tableofcontents*
	\thispagestyle{empty} 
} }

\makeatletter
\newcommand*\ifthispageodd{%
  \checkoddpage
  \ifoddpage
    \expandafter\@firstoftwo
  \else
    \expandafter\@secondoftwo
  \fi
}
\makeatother

\usepackage[noindentafter]{titlesec} 
\counterwithout{section}{chapter} 
\counterwithout{figure}{chapter} 
\counterwithout{table}{chapter} 
\numberwithin{equation}{section} 
\setsecnumdepth{subsubsection}

\DeclareMathVersion{sans}
\SetSymbolFont{operators}{sans}{OT1}{cmbr}{m}{n}
\SetSymbolFont{letters}{sans}{OML}{cmbrm}{m}{it}
\SetSymbolFont{symbols}{sans}{OMS}{cmbrs}{m}{n}
\SetMathAlphabet{\mathit}{sans}{OT1}{cmbr}{m}{sl}
\SetMathAlphabet{\mathbf}{sans}{OT1}{cmbr}{bx}{n}
\SetMathAlphabet{\mathtt}{sans}{OT1}{cmtl}{m}{n}
\SetSymbolFont{largesymbols}{sans}{OMX}{iwona}{m}{n}

\DeclareMathVersion{boldsans}
\SetSymbolFont{operators}{boldsans}{OT1}{cmbr}{b}{n}
\SetSymbolFont{letters}{boldsans}{OML}{cmbrm}{b}{it}
\SetSymbolFont{symbols}{boldsans}{OMS}{cmbrs}{b}{n}
\SetMathAlphabet{\mathit}{boldsans}{OT1}{cmbr}{b}{sl}
\SetMathAlphabet{\mathbf}{boldsans}{OT1}{cmbr}{bx}{n}
\SetMathAlphabet{\mathtt}{boldsans}{OT1}{cmtl}{b}{n}
\SetSymbolFont{largesymbols}{boldsans}{OMX}{iwona}{bx}{n}

\titleformat{\section}{\centering \needspace{4\baselineskip}\Large \bfseries \sffamily \mathversion{boldsans} \color{blue!80!black} }{\thesection}{15pt}{}{}
\titlespacing{\section}{0pt}{15pt}{5pt}
\titleformat{\subsection}{\large \sffamily \mathversion{sans} \color{blue!70!black} }{\thesubsection}{10pt}{}{}
\titlespacing{\subsection}{0pt}{10pt}{5pt}
\titleformat{\subsubsection}{\normalsize \sffamily \itshape \mathversion{sans} \color{blue!70!black} }{\thesubsubsection}{10pt}{}{}
\titlespacing{\subsubsection}{0pt}{10pt}{0pt}

\newcommand{\sectionlike}[1]{\phantomsection \addcontentsline{toc}{section}{#1} \setcounter{subsection}{0} \sectionmark{#1}
		\begin{center}
		\needspace{5\baselineskip}
		\Large \bfseries \sffamily \mathversion{boldsans} \color{blue!80!black} #1  
		\end{center}
	\vspace{-5pt} 
}

\ifnum\fullheadfoot=1
	\makepagestyle{standardstyle}
	\makeoddhead{standardstyle}{\sffamily \mathversion{subsectionmath} \rightmark}{}{\sffamily \bfseries{\thepage}}
	\makeevenhead{standardstyle}{\sffamily \bfseries{\thepage}}{}{\sffamily \mathversion{subsectionmath} \leftmark}
	\makeheadrule{standardstyle}{\textwidth}{\normalrulethickness}
	\makepsmarks{standardstyle}{
		\nouppercaseheads
		\createmark{section}{both}{shownumber}{\ }{\hspace{1.2em}  }
		\createmark{subsection}{right}{shownumber}{}{\hspace{1.2em} }
	}
	
	\copypagestyle{appstyle}{standardstyle}
	\makeevenhead{appstyle}{\sffamily \bfseries{\thepage}}{}{ \sffamily \mathversion{subsectionmath} \leftmark}
	\makepsmarks{appstyle}{
		\nouppercaseheads
		\createmark{section}{both}{nonumber}{\ }{\ \ }
		\createmark{subsection}{right}{shownumber}{}{\ \ }
	}
\else 
	\makepagestyle{standardstyle}
	\makeoddfoot{standardstyle}{}{\sffamily -- {\bfseries\thepage} --}{} 
	\makeevenfoot{standardstyle}{}{\sffamily -- {\bfseries\thepage} --}{}
	\copypagestyle{appstyle}{standardstyle}
\fi

\createplainmark{toc}{both}{\contentsname}
\createplainmark{bib}{both}{\bibname}

\makeatletter
\let\MyIntOrig\int
\def\MyIntSpace{\hspace{-.35em}} 
\def\int{\MyInt}
\def\MyInt{\MyIntOrig\MyIntSkipMaybe}
\def\MyIntSkipMaybe{
	\@ifnextchar_{\MyIntSkipScript}{%
		\@ifnextchar^{\MyIntSkipScript}{%
			\@ifnextchar\limits{\MyIntSkipTok}{%
				\@ifnextchar\nolimits{\MyIntSkipTok}{%
					\MyIntSpace}}}}%
}
\def\MyIntSkipScript#1#2{#1{#2}\MyIntSkipMaybe}
\def\MyIntSkipTok#1{#1\MyIntSkipMaybe}

\newcommand{\pushright}[1]{\ifmeasuring@#1\else\omit\hfill$\displaystyle#1$\fi\ignorespaces}
\makeatother

\usepackage{tikz}
\usepackage[compat=1.1.0]{tikz-feynman}

\usepackage{anyfontsize}

\newcommand{\tr}{\mathop{\mathrm{tr}} }
\newcommand{\eminus}{\vcenter{\hbox{\scalebox{0.6}[1]{$ - $}}}}	

\newcommand{\hc}{+\, \mathrm{h.c.}}

\newcommand{\dd}{\mathop{}\!\mathrm{d}}
\newcommand{\ud}[2]{\phantom{}^{#1}\phantom{}_{#2}}

\newcommand{\sscript}[1]{{\scriptscriptstyle \mathrm{#1}}}

\newcommand{\LL}{\mathrm{L}}
\newcommand{\RR}{\mathrm{R}}

\newcommand{\SU}{\mathrm{SU}}

\newcommand{\muc}{{\bar{\mu}}}
\newcommand{\mud}{{\hat{\mu}}}
\newcommand{\nuc}{{\bar{\nu}}}
\newcommand{\nud}{{\hat{\nu}}}
\newcommand{\rhoc}{{\bar{\rho}}}
\newcommand{\rhod}{{\hat{\rho}}}
\newcommand{\sigmac}{{\bar{\sigma}}}
\newcommand{\sigmad}{{\hat{\sigma}}}

\begin{document}

\thispagestyle{empty}
\renewcommand*{\thefootnote}{\fnsymbol{footnote}}

\begin{center}
    {\sffamily \bfseries \fontsize{19.5}{24}\selectfont \mathversion{boldsans}
    Symmetry Restoration in the SMEFT:\\ Finite Counterterms for a Non-Anticommuting $\gamma_5$\\[-.5em]
    \vspace{.03\textheight}}
    {\sffamily \mathversion{sans} \Large 
    Javier Fuentes-Mart\'in,$^{1}$\footnote{javier.fuentes@ugr.es} 
    Adri\'an Moreno-S\'anchez,$^{1}$\footnote{adri@ugr.es} 
    and \\[5pt]
    Anders Eller Thomsen$^{2}$\footnote{anders.thomsen@unibe.ch}
    }\\[1.25em]
    { \small \sffamily \mathversion{sans} 
        $^{1}\,$Departamento de Física Teórica y del Cosmos, Universidad de Granada,\\
        Campus de Fuentenueva, E–18071 Granada, Spain\\[5pt] 
        $^{2}\,$Albert Einstein Center for Fundamental Physics, Institute for Theoretical Physics, University of Bern, CH-3012 Bern, Switzerland
    }
    \\[.005\textheight]{\itshape \sffamily \today}
    \\[.03\textheight]
\end{center}
\setcounter{footnote}{0}
\renewcommand*{\thefootnote}{\arabic{footnote}}%
\suppressfloats	

\begin{abstract}\vspace{+.01\textheight}
We present a systematic functional approach to calculating symmetry-restoring counterterms in Effective Field Theories (EFTs) regulated using the Breitenlohner--Maison--'t Hooft--Veltman (BMHV) scheme. Building on a recently developed method that employs auxiliary spurion fields, our approach automates the extraction of these counterterms directly from the one-loop effective action, offering a streamlined and efficient procedure. We demonstrate its efficacy by applying it to the Standard Model Effective Field Theory (SMEFT) at dimension six, providing the first complete determination of its symmetry-restoring counterterms in the BMHV scheme. Our work establishes a robust foundation for consistent one-loop matching and two-loop running computations in EFTs with chiral gauge symmetries.
\end{abstract}

\newpage
\section*{Table of Contents}
\toc


\section{Introduction} \label{sec:intro}

One of the most promising avenues to uncovering new physics lies in precision observables: taking our cue from historic, indirect discoveries such as the charm quark and the Standard Model (SM) third family. This program demands not only increasingly refined SM predictions, but also precise beyond-the-SM (BSM) calculations, since many new-physics effects emerge only as higher-order corrections. At the same time, the lack of direct evidence for new states in current experiments suggests that the sought-after new dynamics may reside at energies far beyond the electroweak scale---a hypothesis naturally accommodated in the Effective Field Theory (EFT) framework. In this context, the Standard Model Effective Field Theory (SMEFT) provides a systematic and model‐comprehensive way to parametrize heavy-scale effects and to systematically incorporate loop‐level contributions, laying the theoretical groundwork for the next generation of precision tests. Recent years have witnessed significant advances, both methodological and computational, in precision EFT calculations; see, e.g.,~\cite{Chala:2024llp,LopezMiras:2025gar,Panico:2018hal,Jenkins:2023bls,Fuentes-Martin:2023ljp,Fuentes-Martin:2024agf, Haisch:2024wnw,DiNoi:2024ajj,Born:2024mgz,Aebischer:2025hsx,Naterop:2023dek,Naterop:2024ydo,Naterop:2025lzc,Naterop:2025cwg,Aebischer:2023nnv,Guedes:2024vuf,Fonseca:2025zjb,Aebischer:2025zxg,Misiak:2025xzq,Duhr:2025zqw} and references therein.

In these calculations, loop diagrams inevitably produce ultraviolet divergences that must be regularized and renormalized before yielding physical predictions. Dimensional regularization (DR), where spacetime is analytically continued to $d = 4 - 2\epsilon$ dimensions, remains the premier choice. Yet the presence of chiral fermions forces one to confront the extension of the inherently four-dimensional $\gamma_5$ to $d$ dimensions. In the SMEFT, where many operators involve chiral structures, this $\gamma_5$ scheme dependence becomes particularly relevant; see e.g.~\cite{DiNoi:2023ygk,DiNoi:2025uan,DiNoi:2025uhu,DiNoi:2025arz} for recent discussions. The simplest prescription---naive dimensional regularization (NDR) with a fully anticommuting $\gamma_5$---is algebraically convenient but mathematically inconsistent. It violates cyclicity of Dirac traces and thus requires ``reading-point'' prescriptions~\cite{Korner:1991sx,Kreimer:1993bh} (the determination of which is not always evident~\cite{Bednyakov:2015ooa,Poole:2019txl,Davies:2019onf}); even then, known higher-order anomalies cannot always be fixed without extra, external input~\cite{Chen:2023lus,Chen:2024zju}. As such, the practical utility of NDR in complex EFT settings is limited. By contrast, the Breitenlohner--Maison--'t Hooft--Veltman (BMHV) scheme~\cite{tHooft:1972tcz,Breitenlohner:1977hr} is the only known mathematically sound and self-consistent approach~\cite{Speer:1974cz,Breitenlohner:1975hg,Breitenlohner:1976te, Costa:1977pd, Aoyama:1980yw}. This mathematical rigor, however, comes at the cost of introducing an explicit breaking of chiral symmetries. Such symmetries can---must in the case of gauge symmetries---be reinstated via finite counterterms. Over the past decades, various methods for computing these symmetry-restoring counterterms have been developed~\cite{Martin:1999cc,Sanchez-Ruiz:2002pcf,Belusca-Maito:2020ala,Belusca-Maito:2021lnk,Belusca-Maito:2022wem,Belusca-Maito:2023wah,Stockinger:2023ndm,Cornella:2022hkc}. A particularly elegant strategy consists in introducing auxiliary fields to enforce a spurious gauge invariance~\cite{OlgosoRuiz:2024dzq}, amenable to automation through diagrammatic computer tools like \texttt{Matchmakereft}~\cite{Carmona:2021xtq}. Yet the diagrammatic approach requires pre-defining a basis of possible symmetry-restoring counterterms---a task that becomes increasingly challenging once higher-dimensional operators are considered.

In this work, we combine the auxiliary-field approach of~\cite{OlgosoRuiz:2024dzq} with the functional techniques developed in~\cite{Fuentes-Martin:2024agf}, resulting in a direct method for determining symmetry-restoring counterterms in the BMHV scheme for arbitrary theories---without requiring any prior knowledge of their structure. We implement this approach in a bespoke version of \texttt{Matchete}~\cite{Fuentes-Martin:2022jrf} and derive, for the first time, the full set of dimension-six symmetry-restoring counterterms for the SMEFT at one-loop order. The paper is organized as follows: In Section~\ref{sec:framework}, we review the BMHV scheme and its $d$-dimensional Lagrangian extension. Section~\ref{sec:func_app} presents our functional approach and its implementation in \texttt{Matchete}, while Section~\ref{sec:SMEFTexample} illustrates the procedure with a SMEFT example. The complete set of SMEFT counterterms is provided as supplementary material. We conclude in Section~\ref{sec:conclusions}.

\section{Framework} 
\label{sec:framework}

In the BMHV scheme,\footnote{For a detailed discussion on this scheme, we refer the reader to~\cite{tHooft:1972tcz,Breitenlohner:1977hr,Belusca-Maito:2023wah} and references therein.} $d$-dimensional spacetime is decomposed as a direct sum of a 4-dimensional physical subspace and a $2\epsilon$-dimensional evanescent subspace. Accordingly, Lorentz indices are split as $\mu = (\muc, \mud)$, where \emph{barred} indices ($\muc$) refer to the 4-dimensional components and \emph{hatted} indices ($\mud$) to the $\epsilon$-dimensional ones. This decomposition affects all tensorial quantities. In particular, one introduces two orthogonal metric tensors corresponding to each subspace:
\begin{align} \label{eq:metrics}
g^{\muc\nuc} g_{\muc\nuc} &= 4\,, &
g^{\mud\nud} g_{\mud\nud} &= -2\epsilon\,, &
g^{\mud\nu } g_{\nu \bar{\rho}} &= 0\,.
\end{align}
These metric tensors allow for the projection of $d$-dimensional quantities onto the respective subspaces, simplifying tensor reductions in loop calculations. For example, one finds
\begin{align}
k^\muc k^\nuc = g^{\muc\rho} g^{\nuc\sigma} k_\rho k_\sigma 
= \frac{1}{d} \, g^{\muc\rho} g^{\nuc\sigma} g_{\rho\sigma} k^2 
= \frac{1}{d} \, g^{\muc\nuc} k^2\,,
\end{align}
with analogous expressions holding for contractions involving the $\epsilon$-dimensional subspace.

This decomposition also extends to the Dirac algebra; indeed, it is the defining feature of the BMHV scheme. The $\gamma$ matrices are taken to satisfy the Clifford algebra independently within each subspace: 
\begin{align} \label{eq:gamma_properties}
\{\gamma^{\muc}, \gamma^{\nuc}\} &= -2 g^{\muc\nuc}\,, &
\{\gamma^{\mud}, \gamma^{\nud}\} &= -2 g^{\mud\nud}\,, &
\{\gamma^{\muc}, \gamma^{\nud}\} &= 0\,.
\end{align}
This structure enables a consistent definition of intrinsically 4-dimensional objects, such as the Levi-Civita symbol $\varepsilon_{\muc\nuc\bar\rho\bar\sigma}$ and the $\gamma_5$ matrix. The latter is defined, using the $\varepsilon^{0123}=+1$ convention, as
\begin{align}
\gamma_5 = -\frac{i}{4!} \, \varepsilon_{\muc\nuc\bar\rho\bar\sigma} \gamma^\muc \gamma^\nuc \gamma^{\bar\rho} \gamma^{\bar\sigma}\,,
\end{align}
ensuring that $\gamma_5$ remains strictly 4-dimensional.

This definition leads to the well-known (anti-)commutation relations between $\gamma_5$ and the ordinary Dirac matrices:
\begin{align} \label{eq:g5gmu}
\{\gamma_5, \gamma^{\muc}\} = 0\,, \qquad
[\gamma_5, \gamma^{\mud}] = 0\,.
\end{align}
These relations play a central role in the BMHV formalism and can be shown to be compatible with the cyclicity of the Dirac trace in dimensional regularization to all orders in perturbation theory~\cite{Costa:1977pd,Aoyama:1980yw}.

\subsection{Extending the Lagrangian to $d$ dimensions}
\label{sec:Ext2d}

In dimensional regularization, one must specify how the Lagrangian is extended to $d$ dimensions, including how its four-dimensional symmetries are promoted. While this extension is straightforward in the scalar sector, it introduces important subtleties for gauge fields and (chiral) fermions.

\subsubsection{Bosonic sector}

The presence of chiral interactions in a theory discriminates between physical and evanescent dimensions.\footnote{Chiral projectors are not invariant under the full $d$-dimensional Lorentz group, but instead break it to $\mathrm{O}(1,3) \times \mathrm{O}(d - 4)$.} As a result, there is no symmetry principle requiring all components of $d$-dimensional gauge fields to be treated on equal footing. A convenient and widely adopted prescription is to restrict gauge fields to their four-dimensional components by setting their evanescent parts to zero, $A_\mud = 0$. This effectively confines gauge transformations to the physical subspace, i.e., $\alpha(x_\mu) = \alpha(x_\muc)$, ensuring consistency. This choice defines a particular variant of the BMHV scheme. While alternative prescriptions have been explored in the literature~\cite{OlgosoRuiz:2024dzq,Belusca-Maito:2020ala}, recent works typically adopt this convention for practical computations~\cite{Belusca-Maito:2021lnk,Cornella:2022hkc,Stockinger:2023ndm,Sanchez-Ruiz:2002pcf,Belusca-Maito:2023wah,Ebert:2024xpy,OlgosoRuiz:2024dzq,Kuhler:2025znv}. Importantly, differences between such schemes are evanescent and can be absorbed into finite redefinitions of the physical couplings~\cite{Buras:1989xd,Dugan:1990df,Herrlich:1994kh,Fuentes-Martin:2022vvu}. With this choice, the $d$-dimensional extension of the bosonic kinetic terms becomes 
\begin{align}
\mathcal{L}^{\text{Bos}}_{(d)} = -\frac{1}{4} F^{\muc\nuc} F_{\muc\nuc} - \frac{1}{2} F^{\muc\nud} F_{\muc\nud} + (D_\muc \phi)^\dagger (D^\muc \phi) + (\partial_\mud \phi)^\dagger (\partial^\mud \phi) + \mathcal{L}_{\text{gf}}\,,
\end{align}
where $D_\muc = \partial_\muc - i g A_\muc^a T^a$ is the four-dimensional covariant derivative, $F_{\mu\nu} = [D_\mu, D_\nu]$ (with $ D_\mud = \partial_\mud $) is the $ d $-dimensional field-strength tensor and $\phi$ a generic scalar field. Evanescent terms involving partial derivatives in the $\mud$ directions do not spoil gauge invariance, as we have restricted ourselves to four-dimensional gauge transformations. Thus, the only breaking of gauge symmetry at this stage is restricted to the gauge-fixing terms. Equivalently, one may also write the bosonic kinetic terms compactly in the ordinary form 
\begin{align}
\mathcal{L}^{\text{Bos}}_{(d)} = -\frac{1}{4} F^{\mu\nu} F_{\mu\nu} + (D_\mu \phi)^\dagger (D^\mu \phi) + \mathcal{L}_{\text{gf}}\,,
\end{align}
with the drawback that it obfuscates the details of the $ d $-dimensional extension.

A particularly useful gauge-fixing choice for the BMHV scheme is the so-called \textit{background-field gauge} \cite{Cornella:2022hkc,Abbott:1981ke}, used to compute the \emph{gauge-invariant effective action}. In this gauge, the quantization is performed in such a way that an explicit background gauge invariance—rather than the non-linear BRST transformations—is preserved even after gauge-fixing and ghost terms have been added. In the background-field gauge, the gauge field is decomposed as
\begin{align} \label{eq:BFM}
A_\muc \to A_\muc + \hat{A}_\muc\,,
\end{align}
with $A_\muc$ being the quantum fluctuation and $\hat{A}_\muc$ a background field sourcing the one-particle-irreducible (1PI) Green's functions of the gauge-invariant effective action.\footnote{We emphasize that the hat here denotes background fields and is not related to the evanescent subspace of Lorentz indices.} In this construction, the gauge-fixing term is covariant with respect to the background-gauge transformations. The common choice is to use a background-gauge invariant version of the $ R_\xi $ gauges: 
\begin{align}\label{eq:GaugeFixing}
\mathcal{L}_{\text{gf}} = -\frac{1}{2\xi} (\widehat{D}^\muc A_\muc)^2 + (\widehat{D}_\muc\bar{u}) (D^\muc u) + (\partial_\mud\bar{u})\, (\partial^\mud\, u)\,,
\end{align}
where $\widehat{D}_\muc = \partial_\muc - i g \hat{A}_\muc^a T^a$ is the pure background-field covariant derivative, with the ordinary covariant derivative now being $D_\muc= \partial_\muc - i g (A_\muc^a + \hat{A}_\muc^a) T^a $, and  $u$($\bar u$) is the corresponding (anti-)ghost field, which is purely quantum. In what follows, we adopt this gauge-fixing choice and further set $\xi = 1$ in our calculations.

\subsubsection{Fermionic sector}

The extension of the Lagrangian to $d$ dimensions in the fermionic sector is more involved. The (anti-)commutation relations~\eqref{eq:g5gmu} between \( \gamma^\mu \) and \( \gamma_5 \) imply that
\begin{align}
\gamma^\muc P_{\LL(\RR)} &= P_{\RR(\LL)}\gamma^\muc\,, &
\gamma^\mud P_{\LL(\RR)} &= P_{\LL(\RR)}\gamma^\mud\,,
\end{align}
with the chiral projectors defined as usual: $P_{\LL(\RR)} = \frac{1}{2}(1 \mp \gamma_5)$. 
With the obvious extension to $ d $ dimensions, the four- and $-2\epsilon$-dimensional part of the fermion kinetic operators couple to fermions of different chiralities. 
To regularize fermion propagators in DR, one is required to couple fermions of opposite chirality in the evanescent sector~\cite{Jegerlehner:2000dz}.

Various prescriptions for implementing this extension exist in the literature, each realizing a different BMHV scheme (see~\cite{Ebert:2024xpy} for a recent discussion). For a massive Dirac fermion, $\Psi^i$, a natural choice consists in pairing the different chiralities in the Dirac multiplet; that is,
\begin{align} \label{eq:BMHVDirac}
\mathcal{L}^{\text{Fer}}_{(d)} = i \overline{\Psi}_i \gamma^\muc D_\muc \Psi^i -  \overline{\Psi}_i M\ud{i}{j}\Psi^j + \left(i \overline{\Psi}_i \gamma^\mud P_\LL \partial_\mud \Psi^i \hc \!\right).
\end{align}
The index on the fermion is generic at this stage: it could be gauge, global, flavor, a combination of them, or even cover fermions in reducible representations of the symmetry groups. This extension for Dirac fermions is known to provide simple $d$-dimensional propagators in the massive case; however, the evanescent piece violates the global chiral symmetry, which is broken by the mass term. This prescription can also be applied to chiral fermions by organizing them into faux-Dirac fermions, whose chiralities transform differently under (gauge) symmetries, see, e.g.~\cite{OlgosoRuiz:2024dzq}.\footnote{With this organization, there are no mass terms. The evanescent part of the kinetic term (involving $ \partial_\mud $) connects both chiralities and violates the (chiral) symmetries of the theory, both gauge and global, explicitly.}

For massless fermions—or fermions where the mass is treated as a perturbation—we find it advantageous to follow a different approach, based on the introduction of an \emph{evanescent} chiral partner.\footnote{This is referred to as \emph{sterile} in~\cite{Ebert:2024xpy}, but we adopt the term \emph{evanescent}, which we find more appropriate in this context. The number of evanescent partners must match the number of physical fermions, including the multiplicities due to gauge and global symmetries and flavor.} This fermion partner is made to interact only via the evanescent piece of the kinetic term of its physical counterpart, so it needs to be a singlet under all gauge symmetries. Its only role is to regulate the fermionic kinetic term, and it decouples from the rest of the theory in the four-dimensional limit. 

Let us consider the case of a generic chiral fermion $\psi^i$, taken to be left-handed without loss of generality, with a generic index~$ i $. Under gauge and global symmetries, this fermion transforms as
\begin{align} \label{eq:fermion_transformation}
\psi^i \xlongrightarrow[\,\mathrm{gauge}\,]{} U\ud{i}{j}\, \psi^j\,, \qquad
\psi^i \xlongrightarrow[\,\mathrm{global}\,]{} V\ud{i}{j}\, \psi^j\,.
\end{align}
We associate with this fermion an evanescent right-handed partner $ \psi_{\mathrm{ev}}^i $ with one component for each component in $ \psi^i $. We can think of the index on $ \psi_{\mathrm{ev}}^i $ as some sort of evanescent flavor index, which is a global version of the index on $ \psi^i $.
This fermionic extension lets us construct a $ d $-dimensional kinetic term to regulate the fermion propagator: 
\begin{align} \label{eq:BMHVChiral}
\mathcal{L}^{\text{Fer}}_{(d)} = i \overline{\psi}_i \gamma^\muc P_\LL D_\muc \psi^i + i \overline{\psi}_{\mathrm{ev},i} \gamma^\muc P_\RR \partial_\muc \psi_{\mathrm{ev}}^i + \left( i \overline{\psi}_i\gamma^\mud P_\RR \partial_\mud \psi_{\mathrm{ev}}^{i}  \hc \!\right)\,.
\end{align}
As in the other scheme, the evanescent piece introduces an explicit breaking of chiral gauge symmetries. However, the evanescent fermion $\psi_{\mathrm{ev}}^i$ can be taken to transform under global symmetries like its physical counterpart, ensuring that the extended Lagrangian respects all global symmetries of the original theory (including the chiral ones). We have 
\begin{align}\label{eq:ev_fermion_transformation}
\psi^i_\mathrm{ev} \xlongrightarrow[\,\mathrm{gauge}\,]{} \psi^i_\mathrm{ev}\,, \qquad 
\psi^i_\mathrm{ev} \xlongrightarrow[\,\mathrm{global}\,]{} V\ud{i}{j}\, \psi^j_\mathrm{ev}\,.
\end{align}
The evanescent fermion cannot be allowed to share the gauge transformation of the physical fermion; otherwise, it would not decouple in the four-dimensional limit, and the physical theory would be modified. This is why the regularization in the BMHV scheme invariably violates gauge symmetries. 
In practice, since this approach preserves global symmetries, it typically results in simpler symmetry-restoring counterterms~\cite{Ebert:2024xpy} and has been adopted in recent calculations~\cite{Kuhler:2025znv,Stockinger:2023ndm,vonManteuffel:2025swv}. 

In summary, the Dirac-pairing approach is simpler for massive fermions but explicitly breaks global symmetries, whereas the evanescent-partner prescription, better suited to chiral theories such as the SMEFT, preserves them and enables a more straightforward fermion pairing. We therefore adopt the evanescent-partner prescription in what follows and relegate details on the Dirac-pairing approach to Appendix~\ref{app:Dirac}.

\subsection{Symmetry restoration}
\label{sec:sym-rest}

Any extension of the fermion Lagrangian to $d$ dimensions in the BMHV scheme introduces a breaking of gauge symmetry, which manifests through violations of the Ward–Takahashi identities.\footnote{With other gauge-fixing choices, such as the $R_\xi$ gauge, gauge symmetry is already broken by the gauge fixing itself, and the relevant symmetry is the BRST symmetry. Its breaking leads to violations of the more involved Slavnov–Taylor identities~\cite{Breitenlohner:1977hr,Belusca-Maito:2023wah}.} These symmetry violations originate in the evanescent sector and are thus local; they can be removed by appropriate finite counterterms computed order by order in perturbation theory~\cite{Belusca-Maito:2021lnk,Martin:1999cc,Grassi:1999tp,Cornella:2022hkc,Belusca-Maito:2020ala}. Various strategies have been developed to determine these counterterms. A common approach computes the explicit gauge (or BRST) variation of the regularized action and reconstructs the necessary counterterms by enforcing symmetry restoration~\cite{Cornella:2022hkc,Belusca-Maito:2020ala,Belusca-Maito:2021lnk,Belusca-Maito:2023wah,Ebert:2024xpy,Kuhler:2025znv,vonManteuffel:2025swv}. Another class of methods employs spurion analysis to track the spurious breaking of global chiral symmetries and infer the required operator structures~\cite{Naterop:2023dek,Naterop:2024ydo,Naterop:2025lzc,Naterop:2025cwg}. These techniques, however, are often laborious and difficult to automate.

A recent proposal~\cite{OlgosoRuiz:2024dzq} provides a more streamlined method, directly applicable to EFTs and well suited for automation. It is based on auxiliary spurion (background) fields $\Omega\ud{i}{j}$, which are added to the evanescent part of the fermion kinetic term to restore gauge invariance in the extended $d$-dimensional Lagrangian. Accordingly, one replaces~\eqref{eq:BMHVChiral} with
\begin{align} \label{eq:spurion-lagrangian}
\mathcal{L}^{\mathrm{Fer}}_{(d)} = i\overline\psi_i \gamma^\muc P_\LL D_\muc \psi^i + i\overline\psi_{\mathrm{ev},i} \gamma^\muc P_\RR \partial_\muc \psi_{\mathrm{ev}}^i + \left(i\overline\psi_i \Omega\ud{i}{j} \gamma^\mud P_\RR \partial_\mud \psi_{\mathrm{ev}}^j \hc\right)\,,
\end{align}
and analogously for~\eqref{eq:BMHVDirac}. Like the gauge fields, the spurion fields can, without loss of generality, be restricted to the physical subspace, that is, $\Omega(x_\mu)=\Omega(x_\muc)$. Consequently, derivatives along the evanescent directions vanish, ($\partial_\mud\Omega)=0$, thus providing useful simplifications in intermediate steps. Noting that the physical and evanescent fermions transform as in in~\eqref{eq:fermion_transformation} and~\eqref{eq:ev_fermion_transformation}, respectively, the spurion is chosen to transforms as
\begin{align}
\Omega\ud{i}{j} \xlongrightarrow[\,\mathrm{gauge}\,]{} U\ud{i}{k}\, \Omega\ud{k}{j}\,, \qquad 
\Omega\ud{i}{j} \xlongrightarrow[\,\mathrm{global}\,]{} V\ud{i}{k}\, \Omega\ud{k}{\ell}\, (V^\dagger)\ud{\ell}{j}\,,
\end{align}
so that, by construction, it compensates the transformation mismatch between the physical and evanescent fermions. Since the original Lagrangian is already invariant under global symmetries, it is often convenient to let the spurion transform trivially under them, i.e., to set it proportional to the identity in global (and flavor) indices. In practice, this allows one to drop global and flavor indices in the spurion when using the evanescent-partner scheme.\footnote{For instance, a convenient spurion decomposition for the left-handed SM fields, $\psi^i= \big(q^{ai}_p,\, \ell^{i}_p \big) $ where $a$, $i$ and $p$ denote color, $\SU(2)_L$, and flavor indices, respectively, is 
    \begin{align*}
    \Omega\ud{i}{j} = \begin{pmatrix}
        (\Omega_q)\ud{ai}{bj} \delta_{pr} & \\ & (\Omega_\ell)\ud{i}{j} \delta_{pr}
    \end{pmatrix}\,,
    \end{align*}
which minimizes the spurion degrees of freedom while preserving the required symmetry transformations.
} 
For the scheme in~\eqref{eq:BMHVDirac}, by contrast, $\Omega$ must transform as $\Psi_\LL$ (from the left) and $\overline{\Psi}_\RR$ (from the right) under both gauge and global symmetries. In this case, global indices must be retained, while flavor indices may still be dropped. For practical computations, it is further necessary to impose that $\Omega$ is unitary, i.e., $\Omega \Omega^\dagger = \Omega^\dagger \Omega = \mathds{1} $, so that the denominators of $d$-dimensional loop propagators retain the standard form in DR.

In the \emph{identity limit}, $\Omega \to \mathds{1}$, the original $d$-dimensional Lagrangians~(\ref{eq:BMHVDirac}, \ref{eq:BMHVChiral}) are recovered and the explicit gauge-symmetry breaking reappears. However, since the Lagrangian~\eqref{eq:spurion-lagrangian} is gauge invariant, the resulting quantum effective action is gauge invariant at all loop orders. This property enables a direct prescription to extract the symmetry-restoring counterterms~\cite{OlgosoRuiz:2024dzq}:\footnote{
Briefly summarized, the effective action in the identity limit, $\lim_{\Omega\to\mathds{1}}\Gamma$, coincides with that of the original $d$-dimensional Lagrangian~(\ref{eq:BMHVDirac}, \ref{eq:BMHVChiral}), i.e., before adding symmetry-restoring counterterms. The spurion-free part, $\Gamma_{\slashed\Omega}\equiv\Gamma-\Gamma_\Omega$, trivially preserves all symmetries, as symmetry transformations do not mix terms with and without $\Omega$, whereas $\lim_{\Omega\to\mathds{1}}\Gamma_\Omega$ does not. Hence, the symmetry variations satisfy
$\delta(\lim_{\Omega\to\mathds{1}}\Gamma)=\lim_{\Omega\to\mathds{1}}\delta\Gamma_\Omega$, reproducing the symmetry breaking of the unmodified theory. The counterterms~\eqref{eq:master-formula} are then constructed to cancel this variation, thereby restoring the 
symmetry.
}
    \begin{align} \label{eq:master-formula}
    \Delta S^{(\ell)}_{\mathrm{ct}} = -\lim_{\Omega \to \mathds{1}} \Gamma^{(\ell)}_\Omega\,,
    \end{align}
where $\Delta S^{(\ell)}_{\mathrm{ct}}$ denotes the finite counterterms restoring gauge symmetry, and $\Gamma^{(\ell)}_\Omega$ is the spurion-dependent part of the effective action, both at loop order $\ell$. As all $\Omega$-interactions are evanescent, $\Gamma^{(\ell)}_\Omega$ is local and can be extracted from the UV-divergent part of loop integrals associated with 1PI diagrams involving at least one external spurion field.\footnote{As spurions have no kinetic terms, they do not propagate and can appear only as external (background) fields.} 
The method straightforwardly allows also for the extraction of counterterms needed to restore global symmetries, such as those broken by the Dirac-pairing regularization scheme, cf.~\eqref{eq:BMHVDirac}. For global symmetries, additional simplifications can be obtained by using a constant  $ \Omega $~\cite{OlgosoRuiz:2024dzq,DiNoi:2025uan}. In chiral gauge theories, such as the SMEFT, we use the spacetime-dependent $ \Omega $.

Despite the formal simplicity of~\eqref{eq:master-formula}, implementing it via diagrammatic or amplitude methods remains challenging. The core difficulty lies in incorporating spurion terms into the fermion propagator to properly regulate loop amplitudes. Although one can, in principle, compute the fermion propagator in a background of constant spurions and treat their spacetime dependence perturbatively, the authors of~\cite{OlgosoRuiz:2024dzq} adopted a different approach: they kept the spurions constant throughout the calculation and later reconstructed the effects of spacetime-dependent spurions by projecting onto a pre-defined operator basis. While this strategy simplifies the amplitude calculations, it comes at the cost of requiring a full classification of spurion-dependent operators, a non-trivial task in EFTs.

As we show in the next section, the functional evaluation of~\eqref{eq:master-formula} offers a more efficient and conceptually transparent alternative. It reduces the number of required computations and allows $\Omega$ to be treated as a genuinely spacetime-dependent field, thus avoiding the need to construct operator bases or to expand around static backgrounds.

\section{Functional Evaluation of Symmetry-Restoring Counterterms}
\label{sec:func_app}

In this section, we develop the functional methods used to compute the symmetry-restoring counterterms. We focus on their one-loop determination and postpone the analysis of higher-loop corrections to future work. Modern functional techniques for evaluating the effective action have been extensively discussed in the literature; see~\cite{Henning:2014wua,Fuentes-Martin:2016uol,Zhang:2016pja,Cohen:2020fcu} and references therein. Their extension to multi-loop computations has also been explored in~\cite{Fuentes-Martin:2023ljp,Fuentes-Martin:2024agf,Born:2024mgz}. We provide a brief overview of these methods in the next section and refer the interested reader to the aforementioned works for a more detailed exposition of the formalism and its applications.

\subsection{Functional evaluation of the one-loop effective action}

The functional formalism provides methods for determining the (quantum) effective action, the generating functional for 1PI Green’s functions. It is obtained perturbatively via a Gaussian approximation of the path integral around the background fields.\footnote{In the path integral evaluation, the background-field method is used not only for the gauge fields, as in~\eqref{eq:BFM}, but also for all matter fields of the theory.} The one-loop effective action $\Gamma^{(1)}$ is identified with the supertrace of the logarithm of the \emph{fluctuation operator} $\mathcal{Q}$, defined as the second functional derivative of the tree-level action:
\begin{align}\label{eq:functional-logarithm-Q}
\Gamma^{(1)}=\frac{i}{2}\,\mathrm{STr}\log \mathcal{Q}\,,\qquad \text{where} \quad \mathcal{Q}_{IJ}=\zeta_{JJ}\,\frac{\delta S^{(0)}[\eta]}{\delta \eta_I \delta\eta_J}\Bigg|_{\eta=\hat\eta}\,.
\end{align}
Here, $S^{(0)}[\eta]$ is the tree-level action, and $\eta_I=\eta_a(x)$ denotes the set of \emph{all} fields in the theory in DeWitt notation, with $a$ being an internal index (including Lorentz, flavor, and gauge and global symmetry indices). Repeated indices imply both a sum over internal degrees of freedom and spacetime integration. The factor $\zeta_{IJ}$ is a Grassmannian sign associated with the bosonic or fermionic nature of the fields, taking the value $-1$ when both fields are Grassmannian (fermions and ghosts) and $+1$ otherwise. The indices of $\zeta_{IJ}$ do not count as repeated indices when determining summation. 

The standard procedure for computing the one-loop effective action consists of splitting the fluctuation operator into two parts:
\begin{equation} \label{eq:Q_split}
\mathcal{Q}_{IJ} = \boldsymbol{\Delta}_{IJ} - \mathcal{X}_{IJ}\,,
\end{equation}
where $\boldsymbol{\Delta}$ is identified with the kinetic operator of the fields (but with covariant derivatives), while $\mathcal{X}$ collects the remaining terms, typically referred to as \textit{interaction terms}. After suitable manipulations of the supertrace, the one-loop effective action can be written as a sum of \emph{log-type} and \emph{power-type} contributions:
\begin{align} \label{eq:OneLoopEffectiveAction}
\Gamma^{(1)} = \dfrac{i}{2}\mathrm{STr}\log \boldsymbol{\Delta} - \dfrac{i}{2} \sum_{n=1}^{\infty} \dfrac{1}{n}\, \mathrm{STr} \big(\boldsymbol{\Delta}^{\eminus 1} \mathcal{X} \big)^n \equiv \Gamma^{(1)}_\mathrm{log} + \Gamma^{(1)}_\mathrm{power}\,.
\end{align}
These supertraces can be cast as~\cite{Fuentes-Martin:2024agf}:
\begin{align}
\begin{aligned}\label{eq:supertraceEvaluations}
\mathrm{STr}\log \boldsymbol{\Delta} &=\int \dd^d x \! \int \frac{\dd^dk}{(2\pi)^d} \,  [\log \Delta(x,P_x+k)]_{ab}\,{U^{b}}_a(x,y)\big|_{x=y}\,,\\
\mathrm{STr}\big(\boldsymbol{\Delta}^{\eminus 1} \mathcal{X} \big)^n &=\int \dd^d x \! \int \frac{\dd^dk}{(2\pi)^d} \,  \big\{\big[\Delta^{ \eminus 1}(x,P_x+k) \,X(x,P_x+k) \big]^n \big\}_{ab}\,{U^{b}}_a(x,y)\big|_{x=y}\,.
\end{aligned}
\end{align}
Here, the integral over $k$ arises from the regularization of the supertrace and plays a role analogous to that of the loop-momentum in diagrammatic computations. In these expressions, we utilized that $\boldsymbol{\Delta}$ and $\mathcal{X}$ are local differential operators and take the generic form
\begin{align}
\boldsymbol{\Delta}_{IJ}=\Delta_{ac}(x,P_x)\,{\delta^c}_b(x,y)\,,\qquad 
\mathcal{X}_{IJ}=X_{ac}(x,P_x)\,{\delta^c}_b(x,y)\,,
\end{align}
where $P_x = iD_x$ is the covariant momentum operator and ${\delta^a}_b(x,y) = {U^a}_b(x,y)\,\delta(x-y)$ defines the covariant delta function. The matrix ${U^a}_b(x,y)$ is a Wilson line that parallel transports field representations from point $x$ to point $y$ along the geodesic. As discussed in detail in~\cite{Fuentes-Martin:2024agf} (see also~\cite{Barvinsky:1985an,Kuzenko:2003eb}), the covariant delta function guarantees that all quantities remain gauge covariant throughout the functional evaluation. In particular, the action of covariant derivatives on a Wilson line gives:
\begin{align}
P_x^\nu\, U(x,y)=\sum_{n=1}^\infty\frac{(-1)^{n+1}}{(n+1)!}\,(x-y)_{\mu_1}\cdots (x-y)_{\mu_n} [D_x^{\mu_1}\cdots D_x^{\mu_{n-1}} G^{\mu_n\nu}(x)]\,U(x,y)\,,
\end{align}
which, upon repeated application, yields gauge-covariant expressions in the coincidence limit $x=y$. Closed-form expressions for repeated applications can be found in~\cite{Fuentes-Martin:2024agf}.

The effective action in~\eqref{eq:OneLoopEffectiveAction} is generally non-local, which can be traced to the inverse and logarithm of $\Delta(x,P_x+k)$ being non-local operators, and their evaluation can be quite challenging. As discussed in~\cite{Fuentes-Martin:2016uol,Zhang:2016pja,Fuentes-Martin:2024agf,Born:2024mgz} and references therein, local contributions, such as the matching corrections to an EFT or UV divergent contributions, can be easily extracted from the effective action using the method of regions in DR~\cite{Beneke:1997zp,Jantzen:2011nz}. As we discuss next, the symmetry-restoring counterterms, as any other finite counterterms associated with evanescent contributions, can be extracted using similar techniques~\cite{Fuentes-Martin:2022vvu}.

\subsection{Symmetry-restoring counterterms}

Since the $\Omega$ field is introduced only in evanescent interactions (formally of rank $\epsilon$) they appear in the physical sector only when multiplied by UV poles in loop integrals. A practical way to extract such contributions, at one-loop order, from the effective action is to expand all loop integrands in the \emph{hard} region, defined by $k \gg p_i, m_i$, and formally replace scaleless integrals according to 
\begin{align}
\int\frac{\dd^d k}{(2\pi)^d}\, \frac{1}{(k^2)^n} \longrightarrow \begin{dcases}
    \frac{i}{16\pi^2}\,\frac{1}{\epsilon} & \text{for}\; n=2\\
    0 &\text{for}\; n\neq 2
\end{dcases}\,,
\end{align}
This renders all expressions local and straightforward to evaluate. The symmetry-restoring counterterms are obtained directly from the spurion-dependent terms as per~\eqref{eq:master-formula}. 

When computing the BMHV counterterms, a key difference with respect to standard functional evaluations is that the evanescent kinetic term containing the spurion must be included in the definition of $\boldsymbol{\Delta}$ to properly regularize the loop integrals. Applied to the Lagrangian in~\eqref{eq:spurion-lagrangian}, and arranging the fields as $\chi=(P_\LL\,\psi\;\;P_\RR\,\psi^{\sscript{ev}})$, this leads to the fermion kinetic operator
\begin{align}\label{eq:Delta-matrix}
\Delta_{\overline{\chi}\chi}(x,P_\mu)=
\begin{pmatrix}
\gamma^\muc & \gamma^\mud  \Omega  \\[5pt]
\gamma^\mud \Omega^\dag  & \gamma^\muc
\end{pmatrix}
P_\mu\,,
\end{align}
where the subindex indicates, in order, the arguments of the functional derivative in~\eqref{eq:functional-logarithm-Q}. The choice of working with barred instead of charged conjugated fermion fields typically yields simpler expressions at one loop, which makes it the standard choice. Unlike schemes with an anticommuting $ \gamma_5$, where $\Delta$ in this form is diagonal, we now encounter a non-diagonal structure induced by the mixing between physical and evanescent fermions via the evanescent interaction. Expressions for the hard-region expansion of $\log\Delta_{\overline{\chi}\chi}(x,P_x+k)$ and $\Delta^{\eminus 1}_{\overline{\chi}\chi}(x,P_x+k)$—the basic ingredients of the functional supertrace—are provided in Appendix~\ref{app:expresion-Delta}. As shown there, no constant limit needs to be assumed, and the symmetry-restoring counterterms arise \textit{directly} from the functional computation of the effective action (just as EFT operators appear directly when doing functional matching calculations).

Regarding the interaction terms, our scheme—where evanescent fermions appear only in the evanescent kinetic terms and the gauge symmetry is kept strictly four-dimensional—ensures that evanescent fermions are absent from the $X$ terms. They appear solely in the functional propagator $\Delta$. Consequently, $X$ effectively acts as a projector onto the subspace of physical fields. This feature constitutes a key advantage of the chosen prescription for the regularized fermion Lagrangian, as it implies that only the physical block of $\Delta_{\overline{\chi}\chi}(x,P_x+k)^{-1}$ needs to be considered when evaluating the power-type supertrace.

\subsection{Implementation in \texttt{Matchete}}

The formalism discussed in the previous section has been implemented in a private version of the publicly available \texttt{Mathematica} package \texttt{Matchete}~\cite{Fuentes-Martin:2022jrf}. This extension automates the computation of symmetry-restoring counterterms in the BMHV scheme. In practice, we have made the following changes and additions to our custom version of \texttt{Matchete} to accommodate the BMHV scheme:
\begin{itemize}
    \item We included a complete implementation of the Dirac algebra in the BMHV scheme, with a separation of all Lorentz indices into four- and $ -2\epsilon$-dimensional parts.

    \item The treatment of chiral fermions has been updated according to the prescription in~\eqref{eq:spurion-lagrangian}. In particular, the functional propagator of chiral fermions, $\Delta_{\overline{\chi}\chi}^{\eminus1}$, has been modified to incorporate the corresponding spurion fields, cf.~\eqref{eq:Delta-Power-terms}. These changes affect the implementation of the power-type supertraces in~\eqref{eq:OneLoopEffectiveAction}. The log-type supertrace has likewise been adapted to include spurion contributions, as shown in~\eqref{eq:Delta-Log-terms}.
    
    \item The existing simplification routines in \texttt{Matchete} have been extended to handle spurion-dependent structures, including the enforcement of the unitarity relations $\Omega\Omega^\dagger = \Omega^\dagger\Omega = \mathds{1}$. Additional algebraic identities specific to the BMHV scheme have also been incorporated, most notably the \emph{Chisholm identity}:
    \begin{align}
    \gamma^\muc\gamma^\nuc\gamma^{\bar\rho} = \gamma^\muc g^{\nuc\bar\rho} - \gamma^\nuc g^{\muc\bar\rho} + \gamma^{\bar\rho} g^{\muc\nuc} + i \varepsilon^{\muc\nuc\bar\rho\bar\sigma} \gamma_{\bar\sigma} \gamma_5  \,.
    \end{align}
    Any further simplification in \texttt{Matchete} is performed under the assumption that the relevant quantities are strictly four-dimensional---an approximation valid for one-loop counterterms but one that will need to be reassessed for higher-order corrections.
    
    \item We have implemented a dedicated function to take the identity limit of the spurion field (i.e., $\Omega \to \mathds{1}$) and expand out the corresponding covariant derivatives, as required for extracting the physical counterterms (cf.~\eqref{eq:master-formula}).
\end{itemize}

Although our computations focus on determining BMHV counterterms for the SMEFT, our \texttt{Matchete} extension has been implemented in a general form and can be readily applied to other models. This generality enables validation against several established results in the literature. The spurion implementation has been cross-checked against~\cite{OlgosoRuiz:2024dzq}, which provides the spurion-dependent terms of the effective action for a general renormalizable theory.\footnote{We thank Pablo Olgoso for adapting their BMHV-scheme choice to ours, cf.~\eqref{eq:BMHVChiral}.} Our implementation of the identity-limit has been verified in the Abelian sector of the Standard Model, and further tests have been performed through partial comparisons with~\cite{Belusca-Maito:2020ala,Kuhler:2025znv,Cornella:2022hkc}. In the latter cases, complete comparisons are hindered by differences in the BMHV implementation. For instance, the analyses in~\cite{Belusca-Maito:2020ala,Kuhler:2025znv} are carried out in non-background-field gauges, where some counterterms require insertions of non-physical sources. By contrast, using the background-field gauge avoids such complications. Similarly, the results of~\cite{Cornella:2022hkc}, which provide BMHV counterterms for a broad class of renormalizable theories with gauge bosons and chiral fermions, rely on a different $d$-dimensional extension of the fermion Lagrangian---analogous to~\eqref{eq:BMHVDirac}---making direct comparison of some counterterms non-trivial.

\section{Symmetry-Restoring Counterterms in the SMEFT}
\label{sec:SMEFTexample}

We have applied our method to the calculation of symmetry-restoring counterterms in the SMEFT. To the best of our knowledge, this is the first determination of such counterterms in the literature, marking an initial milestone toward enabling next-to-leading order (NLO) SMEFT calculations within the BMHV scheme. The full set of counterterms---both in the identity limit and in spurion form---is provided as supplementary material in the form of a PDF file and a \texttt{Mathematica} notebook.
We observe that $B$-violating operators do not generate counterterms at dimension six. This is consistent, as any symmetry-restoring counterterms (at this order) involve at most two fermionic fields in addition to scalar or spurion insertions. With these building blocks alone, it is not possible to construct a $B$-violating operator, and thus no counterterms for $B$-violating SMEFT operators arise at this order. By similar arguments, the Weinberg operator generates spurion-dependent contributions only upon double insertion, leading to symmetry-restoring counterterms starting at dimension six.

\begin{figure}[t]
\centering
\begin{tikzpicture} \pgfsetlinewidth{.8pt} 
\begin{feynman}
  \vertex[dot] (f1) at (-2,0) {};
  \vertex[dot] (f2) at (1,0) {};

  \vertex[dot] (c) at (-0.5,2.6) {};

 \vertex at (-0.5,-0.5) {$\Delta_{\bar f f}$};

 \vertex at (-2.1,1.3) {$\Delta_{WW }$};
 \vertex at (1.1,1.3) {$\Delta_{WW }$};

 \vertex at (-0.5,2.95) {$X_{WW }$};
 \vertex at (-2.25,-0.25) {$X_{W\bar f }$};
 \vertex at (1.5,-0.25) {$X_{fW }$};

  \diagram* {
    (f1) -- [fermion] (f2),

    (c) -- [photon] (f1),
    (c) -- [photon] (f2),
  };
\end{feynman}

\begin{feynman}
  \vertex (f1) at (5,0);
  \vertex (f2) at (9,0);
  \vertex (f3) at (6,0.5);
  \vertex (f4) at (8,0.5);
  
  \vertex[dot] (c) at (7,2.232) {};
  
  \vertex (h1) at (8,2.5);
  \vertex (h2) at (6,2.5);
  
  \vertex (o11) at (7.25,0.5);
  \vertex (o12) at (7.25,-0.2);
  
  \vertex (o21) at (6.75,0.5);
  \vertex (o22) at (6.75,-0.2);  

  \vertex at (3.5,1.25) {$\Large \xrightarrow{\hspace*{1cm}}$ \qquad};
  
  \vertex (labelfermion1) at (5,0.35) {$f$};
  \vertex (labelfermion2) at (9,0.35) {$f$};
  
  \vertex (labelgauge1) at (6,1.3) {$W$};
  \vertex (labelgauge2) at (8,1.3) {$W$};
  
  \vertex (labelscalar1) at (8.25,2.75) {$H$};
  \vertex (labelscalar2) at (5.75,2.75) {$H$};

  \vertex (labelspurion1) at (7.5,0.15) {$\Omega$};
  \vertex (labelspurion2) at (6.4,0.18) {$\Omega^\dagger$};

  \diagram* {
    (f1) -- [fermion] (f3),
    (f3) -- [fermion,edge label=$f^{\sscript{ev}}$] (f4),
    (f4) -- [fermion] (f2),
    
    (f3) -- [photon] (c),
    (f4) -- [photon] (c),
    
    (c) -- [scalar] (h1),
    (c) -- [scalar] (h2),
    
    (o11) -- [scalar, draw=gray, densely dashed] (o12),
    (o21) -- [scalar, draw=gray, densely dashed] (o22),
  };
\end{feynman}
\end{tikzpicture}
\caption{Left: Functional graph representing a contribution to the spurion-dependent part of the quantum effective action. The black dot denotes the effective vertex from $\mathcal{O}_{HW}$. Right: The same functional graph but with the $X$ and $\Delta$ terms replaced by their expressions in~(\ref{eq:ExXterms},\ref{eq:ExDeltaterms}).}
\label{fig:Diagram-OHW}
\end{figure}
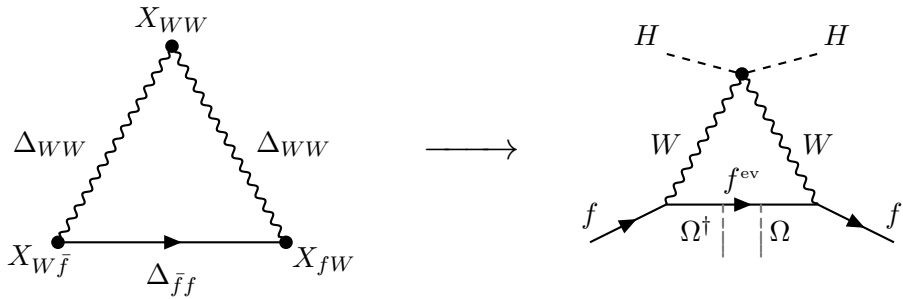

As an illustrative example of our calculations, we consider here the counterterms arising from insertions of the SMEFT operator
    \begin{align}
    \mathcal{O}_{HW} = (H^\dagger H)\,W_{\mu\nu}^I W^{\mu\nu\,I}\,,
    \end{align}
where $I$ is an adjoint $\SU(2)_L$ index. The associated Wilson coefficient is $\mathcal{C}_{HW}$. The only supertrace involving $\mathcal{O}_{HW}$ that contributes to the spurion-dependent part of the effective action is
    \begin{align}    
    \Gamma^{(1)}[\mathcal{O}_{HW}] \supset -i\, \mathrm{STr}\big[\Delta^{\eminus1}_{W_\muc^I W_\nuc^J} X_{W_\nuc^J W_\rhoc^K} \Delta^{\eminus1}_{W_\rhoc^K W_\sigmac^L} X_{W_\sigmac^L f} \Delta^{\eminus1}_{\bar f f} X_{\bar f W_\muc^I} \big]\,,
    \end{align}
where $f = q, \ell$ denotes either SM fermion $\SU(2)_L$-doublet. We have also included a factor of 2 from the complex nature of the fermion fields.\footnote{This degeneracy arises because the operator in question is invariant under charge conjugation. In cases where this invariance does not hold, one must also compute the trace where the fermion fields are replaced with their charge conjugates, noting that the propagator for the conjugated fields coincides with the original one up to the replacement $\Omega \to \Omega^\dagger$.} A graphical representation of this supertrace is shown in the covariant diagram in Figure~\ref{fig:Diagram-OHW} (left). 

The $X$-terms entering the supertrace, derived from the second functional derivatives of $\mathcal{O}_{HW}$ and the fermion kinetic term, are given by
    \begin{align}\label{eq:ExXterms}
    \begin{aligned}
    X_{W_\muc^I W_\nuc^J} &= 4\,\mathcal{C}_{HW}\,\delta_{IJ}\,g^{\muc\rho}g^{\nuc\sigma}\, k_\rho k_\sigma \, (H^\dagger H)  + \mathcal{O}(P)\,,\\
    X_{W_\muc^I f} &= -g_L(\bar f\gamma_\muc| t^I\,, \qquad
    X_{\bar f W_\muc^I} = g_L t^I|\gamma_\muc f)\,,
    \end{aligned}
    \end{align}
where $(\cdot|$ and $|\cdot)$ denote open spinor lines, and $t^I$ are the generators for the fundamental $ \SU(2)_L$ representation. Higher-order terms in the EFT expansion, $\mathcal{O}(P)$, do not contribute at dimension six and are thus neglected. Since the product of $X$-terms is already of EFT dimension five at leading order, the only propagator terms contributing to the spurion-dependent effective action at dimension six are
    \begin{align}\label{eq:ExDeltaterms}
    \Delta_{\bar f f}^{\eminus1} &\supset -\frac{k_\mu k_\nu}{k^4}\,\gamma^\mud \gamma^{\bar\rho} \gamma^\nud \,\Omega_f P_\rho \, \Omega^\dagger_f \,, &
    \Delta_{W_\muc^I W_\nuc^J}^{\eminus1} &= -\frac{1}{k^2}\,g_{\muc\nuc}\,\delta_{IJ} + \mathcal{O}(P)\,,
    \end{align}
having used the expansion in~\eqref{eq:Delta-Power-terms} for the functional fermion propagator, keeping only the leading term with spurion insertions $\Omega_f$. For the gauge propagator, we work in the background-field gauge with $\xi=1$ and set the evanescent components of the gauge fields to zero.

With the $X$-terms and propagators specified, the supertrace can now be evaluated explicitly:
\begin{align}
\Gamma^{(1)}_\Omega[\mathcal{O}_{HW}] &= -4ig_L^2\,\mathcal{C}_{HW}\,t^I_{ki} t^I_{jl}\,(H^\dagger H)\, (\bar f_{kp} \gamma^\muc \gamma^{\rhod} \gamma^{\bar\kappa} \gamma^{\sigmad} \gamma^{\nuc} f_{lp})\, \Omega_f^{i\hat\alpha} (P_{\bar\kappa} \, \Omega_f^{*\,j\hat\alpha}) \int \frac{\dd^d k}{(2\pi)^d} \frac{k^\mu k^\nu k^\rho k^\sigma}{k^8} \nonumber \\
&= -\frac{1}{16\pi^2}\frac{1}{6}\,g_L^2\,\mathcal{C}_{HW}\,(2 \delta\ud{i}{j} \delta\ud{k}{l} - \delta\ud{i}{l} \delta\ud{j}{k})\,(H^\dagger H)\, (\bar f_{ip} \gamma^{\muc} f_p^j) (\Omega_{f,k}^\ast P_{\muc} \, \Omega_f^l)\,,
\end{align}
where we have made explicit the fundamental $\SU(2)_L$ indices ($i,j,k,l$), flavor indices ($p$), and spurion indices ($\alpha$). As discussed in Section~\ref{sec:sym-rest}, the flavor indices have not been included as part of the spurions field. We further use the compact notation where non-flavor indices that are contracted between the fields inside a parenthesis are omitted. In the last line, we used the identity $\Omega^{i\alpha} \Omega^{*\,j\alpha} = \delta_{ij}$ to move the derivative. This result (with $P_{\muc}= i D_{\muc}$) agrees with the expression automatically generated using our modified version of \texttt{Matchete}, provided in Eq.~(3.12) of the supplementary PDF file.

Per~\eqref{eq:master-formula}, the symmetry-restoring counterterms are obtained by taking the identity limit for the spurions. The ones involving the $\SU(2)_L$ gauge fields are 
\begin{align} 
\Delta S^{(1)}_{\mathrm{ct}} \supset -\frac{1}{16\pi^2}\frac{1}{6}\,g_L^3\,\mathcal{C}_{HW}(H^\dagger H)\, (\bar f_p \gamma^{\muc} t^I f_p)\, W_\muc^I\,,
\end{align}
The terms with other gauge fields are analogous to the ones shown here, up to trivial replacements of the generators or gauge charges. This result also agrees with the automatically generated expression available in Eq.~(2.11) of the supplementary PDF file. As expected, the resulting counterterms are not gauge invariant and are therefore only defined up to the addition of gauge-invariant terms, leading to different---but equivalent---renormalization schemes.

\section{Conclusions and Outlook}
\label{sec:conclusions}

In this work, we have combined the auxiliary‐spurion strategy of~\cite{OlgosoRuiz:2024dzq} with modern functional techniques to derive symmetry‐restoring counterterms in the BMHV scheme. Unlike diagrammatic methods, our functional approach naturally incorporates the full spacetime dependence of spurion fields, requires no a priori classification of counterterm operators, and yields the complete counterterm Lagrangian directly from the one‐loop effective action. Crucially, gauge invariance remains manifest at all stages, drastically reducing algebraic complexity and the number of intermediate manipulations. This not only improves computational efficiency but also offers a clearer conceptual picture of the structure of symmetry-restoring counterterms. As a result, our functional approach provides a transparent and automation-friendly framework, ideally suited for EFT calculations.

Building on these methods, we have developed a dedicated version of the \texttt{Matchete} package that fully automates the computation of symmetry-restoring counterterms. Applying this machinery to the SMEFT, we have performed the first complete determination of its symmetry-restoring counterterms, which are provided as ancillary material, both in a human-readable format and as a \texttt{Mathematica} notebook. These results are specifically applicable to SMEFT calculations in the unbroken phase; in the broken phase, where fermions form Dirac fields, the Dirac-pairing scheme (Appendix~\ref{app:Dirac}) is more appropriate, requiring the derivation of new counterterms. Our work thus provides the initial steps toward the more ambitious goal of consistent NLO matching and two‐loop running in chiral gauge theories. 

Looking ahead, we plan to integrate this framework into a future public release of \texttt{Matchete}, thereby providing the community with a tool capable of performing BMHV-consistent one-loop matching and renormalization for generic EFTs. This development is particularly timely, as \texttt{Matchete} is already evolving towards automated two-loop running computations~\cite{Born:2024mgz}. We expect that these results will also be of use for concrete phenomenological studies, such as NLO corrections in the SMEFT for Higgs and precision‐electroweak observables.

\subsection*{Acknowledgments}

We are grateful to Pablo Olgoso and Luiz Vale Silva for valuable discussions and for providing essential cross-checks. The work of JFM and AMS is supported by the grants PID2022-139466NB-C21 funded by MICIU/AEI/10.13039/501100011033 and FEDER/UE, and by the Junta de Andaluc\'ia grants P21\_00199 and FQM101. The work of JMF is further supported by grants EUR2024.153549 and CNS2024-154834 funded by MCIN/AEI/10.13039/ 501100011033 and the European Union NextGenerationEU/PRTR. The work of AMS is further supported by the Spanish government and European Union – NextGenerationEU under grants AST22 6.5 and  FPU23/01639. The work of AET is funded by the Swiss National Science Foundation (SNSF) through the Ambizione grant ``Matching and Running: Improved Precision in the Hunt for New Physics,'' project number 209042.

\renewcommand{\thesection}{\Alph{section}}
\appendix

\section{Expansion of the Spurion-Dependent Propagators} 
\label{app:expresion-Delta}

This appendix collects practical formulae for evaluating power- and log-type supertraces in the presence of spurion fields, used to determine the one-loop symmetry-restoring counterterms. The evanescent-partner~\eqref{eq:BMHVChiral} and Dirac-pairing~\eqref{eq:BMHVDirac} schemes are discussed separately.

\subsection{Evanescent-partner scheme} 

To evaluate the functional supertraces, one must expand both the inverse and the logarithm of the differential operator $\Delta_{\bar\chi\chi}(x,P_\mu+k_\mu)$ associated to the spurion-extended Lagrangian. In the evanescent-partner scheme, the relevant Lagrangian is given in~\eqref{eq:spurion-lagrangian}, while the corresponding differential operator, specified in~\eqref{eq:Delta-matrix}, reads:
\begin{align}\label{eq:Delta-matrix-shift}
\Delta_{\bar\chi\chi}(x,P_\mu+k_\mu)=
    \begin{pmatrix}
        \gamma^\muc & \gamma^\mud  \Omega(x)  \\[1.5 ex]
        \gamma^\mud \Omega^\dag(x)  & \gamma^\muc 
    \end{pmatrix}
(P_\mu+k_\mu)\,.
\end{align}
In what follows, we provide the expressions for the functional propagator, $\Delta_{\bar\chi\chi}^{\eminus1}(x,P_\mu+k_\mu)$, and the functional logarithm, $\ln\Delta_{\bar\chi\chi}(x,P_\mu+k_\mu)$.

\subsubsection{Functional propagator}

To evaluate the functional propagator, it is convenient to split $\Delta_{\bar\chi\chi}(x,P_\mu+k_\mu)$ into two pieces:
\begin{align}
\Delta_{\bar\chi\chi}(x,P_\mu+k_\mu)= A+B = B(\mathds{1}+B^{\eminus 1}A)\,,    
\end{align}
with 
\begin{align}
A=
\begin{pmatrix}
    \gamma^\muc  & \gamma^\mud \Omega \\[1.5 ex]
    \gamma^\mud \Omega^\dag & \gamma^\muc
\end{pmatrix} 
P_\mu\,,\qquad
B=k^2 B^{\eminus1}=
\begin{pmatrix}
    \gamma^\muc  & \gamma^\mud \Omega \\[1.5 ex]
    \gamma^\mud \Omega^\dag & \gamma^\muc       
\end{pmatrix}k_\mu \,.
\end{align}
This separation allows for a simple evaluation of the inverse in terms of $B^{\eminus1}$:
\begin{align}\label{eq:Delta-Power-terms}
\begin{aligned}
\Delta^{\eminus 1}_{\bar\chi\chi} = & \big(\mathds{1}+B^{\eminus 1}A\big)^{\!\eminus 1}B^{\eminus 1}= \sum_{n=0}^\infty(-1)^n \big(B^{\eminus 1}A\big)^nB^{\eminus 1}\\    
= &\sum_{n=0}^\infty\frac{(-1)^n}{k^{2(n+1)}}k_{\nu_1}\cdots k_{\nu_{n+1}} \\
&\phantom{......}\times\prod^n_{i=1}
\begin{pmatrix}             
    (\gamma^{\bar{\nu}_{i}}\gamma^{\muc}+\gamma^{\hat{\nu}_{i}}\gamma^{\mud})P_{\mu} & 
    ( \gamma^{\bar{\nu}_i}\gamma^{\mud}+ \gamma^{\hat{\nu}_i}\gamma^{\muc})\Omega P_{\mu} \\[1.5 ex]
    ( \gamma^{\bar{\nu}_i}\gamma^{\mud}+ \gamma^{\hat{\nu}_i}\gamma^{\muc})\Omega^{\dag}P_{\mu}& (\gamma^{\bar{\nu}_{i}}\gamma^{\muc}+\gamma^{\hat{\nu}_{i}}\gamma^{\mud})P_{\mu}
\end{pmatrix}
\begin{pmatrix}
    \gamma^{\nuc_{n+1}}& 
    \gamma^{\nud_{n+1}} \Omega \\[1.5 ex]
    \gamma^{\nud_{n+1}} \Omega^\dag& \gamma^{\nuc_{n+1}}     
\end{pmatrix}\,.
\end{aligned}
\end{align}
Since $X$-terms exclusively involve physical fields, the evanescent entries of the functional propagator are not needed. In practice, this means that only the top-left entry in the matrix product needs to be retained for the power-type supertrace evaluation. As expected, the denominator in this expression is written in terms of $d$-dimensional loop momenta, thus providing the usual regulation of the loop integrals in DR.

\subsubsection{Functional logarithm}

Fermionic log-traces require a bosonization of the $\Delta$ matrix inside the logarithm. That is,\footnote{Technically, bosonization in this manner is allowed only because the logarithm appears inside a Dirac trace in~\eqref{eq:supertraceEvaluations}. We will omit this trace here.}
\begin{align}\label{eq:expansion_log-trace}
    \log \Delta_{\bar\chi\chi}=\frac{1}{2}\log \Delta_{\bar\chi\chi}^2=\frac{1}{2}\log k^2+\frac{1}{2}\log (\mathds{1}+C)\,,
\end{align}
with the $C$ matrix defined as
\begin{align}
\footnotesize
C=\frac{1}{k^2}\begin{pmatrix}
   (\gamma^\muc \gamma^\nuc+\gamma^\mud\gamma^\nud) P_\mu P_\nu+2k^\mu P_\mu&    (\gamma^\muc\gamma^\nud+\gamma^\mud\gamma^\nuc)\Omega P_\mu P_\nu+\gamma^\muc \gamma^\nud [P_\mu,\Omega\, ](P_\nu+k_\nu)\\[1.5 ex] (\gamma^\muc\gamma^\nud+\gamma^\mud\gamma^\nuc)\Omega^\dag P_\mu P_\nu+\gamma^\muc \gamma^\nud [P_\mu,\Omega^\dag\, ](P_\nu+k_\nu)&
   (\gamma^\muc \gamma^\nuc+\gamma^\mud\gamma^\nud) P_\mu P_\nu+2k^\mu P_\mu
\end{pmatrix}\,.
\end{align}
The first term in~\eqref{eq:expansion_log-trace} cancels against the path-integral normalization and can therefore be ignored. The remaining term can be expanded in loop momenta, yielding the following expression:
\begin{align}\label{eq:Delta-Log-terms}
\begin{aligned}
&\log\Delta_{\bar\chi\chi} =\sum_{n= 1}^\infty \frac{(-1)^{n+1}}{2n}\,C^n
=\sum_{n=1}^\infty \frac{(-1)^{n+1}}{2n}\frac{1}{k^{2n}}\\
& \times \prod_{i=1}^n 
{\footnotesize
\begin{pmatrix}
   (\gamma^\muc \gamma^\nuc+\gamma^\mud\gamma^\nud) P_\mu P_\nu+2k^\mu P_\mu&    (\gamma^\muc\gamma^\nud+\gamma^\mud\gamma^\nuc)\Omega P_\mu P_\nu+\gamma^\muc \gamma^\nud [P_\mu,\Omega\, ](P_\nu+k_\nu)\\[1.5 ex] (\gamma^\muc\gamma^\nud+\gamma^\mud\gamma^\nuc)\Omega^\dag P_\mu P_\nu+\gamma^\muc \gamma^\nud [P_\mu,\Omega^\dag\, ](P_\nu+k_\nu) & 
   (\gamma^\muc \gamma^\nuc+\gamma^\mud\gamma^\nud) P_\mu P_\nu+2k^\mu P_\mu
\end{pmatrix}
}\,,
\end{aligned}
\end{align}
Contrary to the functional propagator, both physical and evanescent entries of this expression are needed for the evaluation of the log-type supertrace. We remind the reader that, in any case, only the terms containing spurions should be retained for the determination of the symmetry-restoring counterterms.

\subsection{Dirac-pairing scheme}
\label{app:Dirac}

Massive Dirac fermions connect the different chiralities through the mass term, cf.~\eqref{eq:BMHVDirac}. In this case, the regularized kinetic term is more conveniently written as
\begin{align}
\mathcal{L}^{\sscript{Fer}}_{(d)}=i\bar\Psi\gamma^\muc D_\muc\Psi+\big(i\bar\Psi_L\gamma^\mud \Omega \,\partial_\mud\Psi_R-\bar\Psi_LM\Psi_R\hc\big)\,,
\end{align}
where we omit indices for simplicity and $\Psi_{L,R}$ represent the different chiralities of the Dirac fermion $\Psi$. The spurion is introduced to preserve chiral global symmetries, which are otherwise broken by the mass term. One can formally restore this symmetry by promoting the mass term, and the corresponding Wilson coefficients, to spurions that transform appropriately under the symmetry, see e.g.~\cite{Naterop:2023dek}. In this case, one needs to impose an extra constraint on the spurion: $\Omega \,M^\dagger=M\, \Omega^\dagger$, to obtain standard propagators in DR. Arranging the field as $\chi=(P_\LL\,\Psi\;\;P_\RR\,\Psi)$, we proceed as we did for chiral theories to get $\Delta_{\bar\chi\chi}(x,P_x+k)$  
\begin{align}
    \Delta_{\bar\chi\chi}(x,P_x+k)=\begin{pmatrix}
        \gamma^\muc(P_\mu+k_\mu) & \gamma^\mud \,\Omega(P_\mu+k_\mu)-M\\[1.5 ex]
        \gamma^\mud \,\Omega^\dagger(P_\mu+k_\mu)- M^\dagger   & \gamma^\muc(P_\mu+k_\mu) 
    \end{pmatrix}.
\end{align}
As before, we provide in what follows the expressions for the functional propagator, $\Delta_{\bar\chi\chi}^{\eminus1}(x,P_\mu+k_\mu)$, and the functional logarithm, $\ln\Delta_{\bar\chi\chi}(x,P_\mu+k_\mu)$.

\subsubsection{Functional propagator}

Once again, it is convenient to split the differential operator as
\begin{align}
\Delta_{\bar\chi\chi}(x,P_\mu+k_\mu) = A + B(M)= B(M)\big[\mathds{1}+B^{\eminus1}(M)A\big]\,,    
\end{align}
where now
\begin{align}
A=
\begin{pmatrix}
    \gamma^\muc  & \gamma^\mud \Omega \\[1.5 ex]
    \gamma^\mud \Omega^\dag & \gamma^\muc
\end{pmatrix} 
P_\mu\,,\qquad
B(M)=(k^2-M^2) B^{\eminus1}(-M)=
\begin{pmatrix}
    \slashed{\overline{k}}  & \slashed{\hat{k}} \Omega - M \\[1.5 ex]
    \slashed{\hat{k}} \Omega^\dag - M^\dagger & \slashed{\overline{k}}    
\end{pmatrix} \,.
\end{align}
where $M^2 \equiv M^\dagger M$ and we use the notation $\slashed{\overline{k}}\equiv \gamma^\muc k_\muc$ and $\slashed{\hat{k}}\equiv \gamma^\mud k_\mud$ for compactness. The evaluation of the inverse proceeds along similar lines as before, yielding
\begin{align}\label{eq:DiracPropagator}
\begin{aligned}
    \Delta_{\bar\chi\chi}^{ \eminus 1} & =   \big[\mathds{1}+ B^{\eminus1}(M) A \big]^{\eminus 1}B^{\eminus 1}(M) = \sum_{n=0}^\infty(-1)^n \big[B^{\eminus 1}(M)A\big]^nB^{\eminus 1}(M)\\    
    = &\sum_{n=0}^\infty\frac{(-1)^n}{(k^2-M^2)^{n+1}}\\  
     &\times\prod^n_{i=1}
     {\small
     \begin{pmatrix}
       (\slashed{\overline{k}}\gamma^{\muc} +\slashed{\hat{k}}\gamma^{\mud}+\gamma^\mud M\Omega^\dagger )P_{\mu} & [( \slashed{\overline{k}}\gamma^{\mud} + \slashed{\hat{k}}\gamma^{\muc})\Omega+\gamma^\muc M] P_{\mu} \\[1.5 ex]
        [(\slashed{\overline{k}}\gamma^{\mud} + \slashed{\hat{k}}\gamma^{\muc})\Omega^\dagger+\gamma^\muc M^\dagger] P_{\mu}& (\slashed{\overline{k}}\gamma^{\muc}+\slashed{\hat{k}}\gamma^{\mud}+\gamma^\mud M\Omega^\dagger )P_{\mu}
    \end{pmatrix}\begin{pmatrix}
        \slashed{\overline{k}}   &  \slashed{\hat{k}} \Omega +M \\[1.5 ex] 
        \slashed{\hat{k}} \Omega^\dagger +M^\dagger & \slashed{\overline{k}}        
    \end{pmatrix}\,.
    }
\end{aligned}
\end{align}
As can be seen, the resulting expression is fully analogous to~\eqref{eq:Delta-Power-terms}, differing only by the presence of the mass term. Unlike in that case, however, all entries of the functional propagator now contribute to the power-type supertrace. In the scheme where two chiral fermions are combined into faux-Dirac fermions, the result is obtained trivially by setting $M^2\to0$---thus recovering the same expression as in~\eqref{eq:Delta-Power-terms}--- and considering all entries of the functional propagator.

\subsubsection{Functional logarithm}

Due the presence of the mass term, the bosonization of the log-trace now requires the involvement of the Dirac trace:
\begin{align}
    \tr\log \Delta_{\bar\chi\chi}=\frac{1}{2}\tr\log \Delta_{\bar\chi\chi}\tilde\Delta_{\bar\chi\chi}=2\log (k^2-M^2)+\frac{1}{2}\tr\log (\mathds{1}+C)\,,
\end{align}
where we have defined $\tilde{\Delta}_{\bar\chi\chi}\equiv\Delta_{\bar\chi\chi}\big|_{M^{(\dagger)}\to -\,M^{(\dagger)}}$ and used the fact that the Dirac algebra ensures the trace depends only on even powers of the mass. From this, one readily finds that, as in the case of the functional propagator, the expression for the Dirac trace of the functional logarithm coincides with that in the evanescent-partner scheme (cf.~\eqref{eq:Delta-Log-terms}), up to the replacement in the denominator $k^2\to k^2 - M^2$.

\sectionlike{References}
\vspace{-10pt}
\bibliography{References} 	

\end{document}